\documentclass[11pt,preprint]{aastex}

\received{16 January 2015}
\accepted{2 February 2015}
\slugcomment{Accepted for publication in the Astrophysical Journal}
\shorttitle{Destruction of IS Dust in Evolving SNR Shocks}
\shortauthors{Slavin, Dwek \& Jones}
\begin{document}

\title{Destruction of Interstellar Dust in Evolving Supernova Remnant Shock
Waves}

\author{Jonathan D. Slavin}
\affil{Harvard-Smithsonian Center for Astrophysics, 60 Garden Street,
Cambridge, MA 02138}
\email{jslavin@cfa.harvard.edu}

\author{Eli Dwek}
\affil{Observational Cosmology Lab., Code 665, NASA at Goddard Space Flight
Center, Greenbelt, MD 20771, USA}

\and

\author{Anthony P. Jones}
\affil{Institut d’Astrophysique Spatiale (IAS), UMR 8617, CNRS/Universit\'e
Paris-Sud, 91405 Orsay, France}

\begin{abstract}
Supernova generated shock waves are responsible for most of the destruction of
dust grains in the interstellar medium (ISM). Calculations of the dust
destruction timescale have so far been carried out using plane parallel steady
shocks, however that approximation breaks down when the destruction timescale
becomes longer than that for the evolution of the supernova remnant (SNR)
shock.  In this paper we present new calculations of grain destruction in
evolving, radiative SNRs. To facilitate comparison with the previous study by
\citet{Jones_etal_1996}, we adopt the same dust properties as in that paper.
We find that the efficiencies of grain destruction are most divergent from
those for a steady shock when the thermal history of a shocked gas parcel in
the SNR differs significantly from that behind a steady shock. This occurs in
shocks with velocities $\gtrsim 200$ km s$^{-1}$ for which the remnant is just
beginning to go radiative.  Assuming SNRs evolve in a warm phase dominated
ISM, we find dust destruction timescales are increased by a factor of $\sim 2$
compared to those of \citet{Jones_etal_1996}, who assumed a hot gas dominated
ISM.  Recent estimates of supernova rates and ISM mass lead to another factor
of $\sim 3$ increase in the destruction timescales, resulting in a silicate
grain destruction timescale of $\sim 2$--3 Gyr.  These increases, while not
able resolve the problem of the discrepant timescales for silicate grain
destruction and creation, are an important step towards understanding the
origin, and evolution of dust in the ISM.
\end{abstract}

\section{Introduction}
The evolution of dust grains in the interstellar medium (ISM) is governed by
the processes of creation, including those that add material to pre-existing
grain cores, and destruction, those processes that return grain material back
to the gas phase.  Grain cores are created by thermal condensation in stellar
sources such as the quiescent outflows of asymptotic giant branch (AGB) stars
or the explosively ejected material in supernova (SN) events and may grow via
cold accretion onto refractory grain cores in the dense ISM.  Grains are
destroyed by sputtering and vaporization in grain-grain collisions in
interstellar shocks.  Different processes are important in different
interstellar environments with dense environments, such as cold \ion{H}{1}
clouds and molecular clouds leading primarily to grain growth while lower
density environments, such as diffuse, warm ($T \sim 5000 - 10^4$ K) either
ionized or neutral gas, are mainly sites for grain destruction.  As has been
recognized for many years, the dominant means of destruction for grains is via
fast shock waves mostly generated in the ISM by supernova explosions.  Grains
are destroyed in shocks by a variety of processes \citep[see e.g.,][hereafter
JTH96]{Jones_etal_1996} including thermal sputtering, non-thermal (or
inertial) sputtering and shattering and vaporization caused by grain-grain
collisions.  Both the compression in the shock, with the accompanying
enhancement of the magnetic field and betatron acceleration of the grains, and
the heating of the gas (for non-radiative shocks) act to enhance the
sputtering of the grains.

Grain destruction in shocks has been the subject of numerous studies over
the past several decades \citep[e.g.,][]{Shull_1977,Draine+Salpeter_1979b,
Seab+Shull_1983,McKee_etal_1987,Jones_etal_1994,Jones_etal_1996,
Dwek_etal_1996,Slavin_etal_2004,Bocchio_etal_2014}.  
In most of these works, the shock profile used was that for a steady, plane
parallel shock \citep[though][have approximated shock
evolution]{Draine+Salpeter_1979b,McKee_etal_1987}.
\citet{Dwek_1981} calculated the time-dependent dust destruction in supernova
remnants, but only during the Sedov-Taylor phase of their evolution. The
rationale for using steady shocks is that the processing timescale is short
compared with the evolution timescale for the shock and that the processing
occurs within a distance that is small enough that the difference between a
plane parallel and spherical shock is negligible. As we show below, neither of
these assumptions is completely justified for supernova remnant (SNR) shocks
in the warm, low density medium, which is the most important phase in the ISM
for grain destruction.

\section{Methods}
To assess the effects of hydrodynamical shock evolution on grain destruction,
we have carried out numerical hydrodynamical calculations of supernova remnant
(SNR) evolution.  We have then used that data to construct shock profiles for
many gas parcels for which we have then used our grain processing codes to
calculate the evolution of dust contained in those parcels.

\subsection{Hydrodynamical Calculations}
Our hydrodynamical calculations presented here were done with our own code,
which is based on (and shares some code with) the VH-1 code
(\url{http://astro.physics.ncsu.edu/pub/VH-1/}), which
employs a slightly simplified version of the piece-wise parabolic method of
\citet{Colella+Woodward_1984}.  Ours is a 1D code, assuming spherical
symmetry and includes several enhancements necessary for our calculations.
It uses the same method as VH-1, taking a Lagrangian step, followed by remap
onto a fixed grid.  The enhancements relative to the VH-1 code include:
optically thin radiative cooling, non-equilibrium ionization, thermal
conduction and magnetic pressure.  

\subsubsection{Radiative Cooling and Ionization}
Radiative cooling is included using a set of look-up tables with cooling rate
per ion vs.\ temperature that were generated using an updated version of the
Raymond \& Smith code \citep{Raymond+Smith_1977}.  The cooling rate uses the
current ionization state, calculated as described below, and is thus
consistent with the ionization, which leads to a large enhancement of the
cooling relative to that for collisional ionization equilibrium in the
immediate post-shock region where the gas is underionized.  The elements
included in the ionization and cooling rate calculations are H, He, C, N, O,
Ne, Mg, Si, S, Ar, Ca, Fe, and Ni.  The abundances assumed are those of
\citet{Grevesse_1984} in order to be in keeping with the plane parallel shock
calculations used by JTH96 with which we will compare results. We do not
include dust cooling, which affects the pressure in the hot bubble at late
times \citep{Dwek_1981}. To maintain the temperature in the ambient medium and
prevent cooling to temperatures well below $10^4$ K in the post-shock region,
we include a heating rate that balances the collisional ionization equilibrium
cooling rate at $T = 10^4$ K.  The heating is turned on only for $T < 10^4$ K
and cooling is turned off for those temperatures.  We do not expect this to
have any significant effects on the grain destruction because in the cold
shell the compression is governed by the magnetic field and thermal sputtering
is insignificant at $T \lesssim 10^4$ K. 

The non-equilibrium ionization calculations use collisional ionization and
recombination rates from look-up tables generated similarly to the cooling
rate tables.  We also include a photoionization rate for H and He as we
discuss below. The time advanced ionization states for all the ions considered
are found using a simple tridiagonal matrix inversion of the ionization
equations for each element, using explicit, first order time differencing,
with the exception of H and He for which the ionization equations can be
solved in a simple closed form (given constant rates).  The solution to the
explicit finite difference equations becomes unreliable if the timestep is
larger than $\Delta t \gtrsim \min(1/(\alpha + \beta))$, where $\alpha$ and
$\beta$ are the recombination and ionization rates respectively (including the
electron density) and $\min$ indicates the minimum of the quantity for the
given element.  As a result we divide up the timesteps to subdivisions of the
hydrodynamical timestep so as to not violate the ionization timestep
constraint.  This is the same method we have used previously and it has been
tested and proven accurate \citep{Slavin+Cox_1992,Slavin+Cox_1993}.  We update
the ionization states of the ions after the Lagrangian step and before the
remap since the number of each ion in a parcel should be conserved during the
Lagrangian step.  The fluxes of the ions must be accounted for during the
remap and we take care to ensure that the sum of the ion fractions is one
before and after the remap.

Another influence on the ionization and cooling in the shock is the
pre-ionization of the gas entering the shock front.  Modelers of plane
parallel shock fronts have dealt with this question in a variety of ways.
\citet{Shull+McKee_1979} calculated the ionization created by the radiative
precursor, finding self-consistent values for the ionization in the ambient
medium.  Self-consistent in this context means that the radiative precursor
creates an ionization level that is the same as that presumed when calculating
the radiation field created by the shock.  \citet{Cox+Raymond_1985} took a
different approach.  Arguing that in most cases the upstream gas does not have
time to reach ionization equilibrium, they calculated ``families'' of
pre-ionization-dependent radiative shocks in which self-consistency was not
assumed.  They were primarily concerned with shock diagnostics and found that
faster shocks going into a low pre-ionization medium were very similar to
slower shocks going into a medium with higher pre-ionization. For our
purposes, we are mainly concerned with the post-shock cooling and how it
affects the dynamics and temperature in the shock.  Given the computational
cost in calculating the ionizing radiation field for each timestep and the
fact that an assumption of ionization equilibrium in the pre-shock medium is
probably not justified in most cases, we instead take the simple approach of
assuming either a relatively high ionization level, $X(\mathrm{H}^+) =
X(\mathrm{He}^+) = 0.5$ or a fairly low ionization level $X(\mathrm{H}^+) =
X(\mathrm{He}^+) = 0.1$ for the ambient medium.  With our non-equilibrium
ionization scheme, this requires fixing a photoionization rate necessary to
achieve this level of ionization which then applies everywhere and at all
times during the run.  We also need to fix a heating rate in the medium
necessary to maintain the ambient temperature assumed, $10^4$ K. We have found
that the resulting SNR evolution is not strongly affected by the different
values for the ionization level and the efficiency of grain destruction (the
fraction of grain mass returned to the gas phase) as a function of shock speed
is essentially the same, deviating by at most 6\% (see Table
\ref{tab:mass_dest}). In the rest of this paper we assume a 50\%
pre-ionization in the medium unless stated otherwise.

\subsubsection{Thermal Conduction}
We also include electron thermal conduction in our code using Spitzer
conductivity limited by saturation \citep{Cowie+McKee_1977}.  This is done in
an operator split way and using a two step predictor-corrector approach.  The
first step is a fully implicit half timestep using the initial temperatures to
calculate the conductivities.  The second step uses the temperatures from the
half timestep for calculating the conductivities and then does a full
timestep, this time using a Crank-Nicholson \citep{Crank+Nicolson_1996}
timestep. For both steps the time-advanced temperature is found via
tridiagonal matrix inversion. This approach has proven stable and accurate.
Magnetic fields severely inhibit electron thermal conduction except along the
field lines. We do not include any reduction of the conduction caused by the
magnetic field, but we feel this is justified since conduction is only
important in the hot gas where the field should be dynamically unimportant.
\citet{Balsara_etal_2008} have examined SNR evolution with anisotropic thermal
conduction and found that the anisotropy can be significant, though primarily
at later times in the evolution.  We assume equal electron and ion
temperatures in this work.

\subsubsection{The Magnetic Field}
The magnetic field is included approximately via a pressure term proportional
to the density.  This treatment of the field only applies to the perpendicular
component of the magnetic field and ignores the magnetic tension, which would
require at least a 2D calculation.  Carrying out the calculation in 1D has
allowed us to go to very high resolution, 0.0125 pc, over a volume that
extends to 75 pc and is necessary in the post-shock region of the
radiative shock.  In addition, we are not aware of any publicly available MHD
codes that include all of the physical effects that we have included in our
calculations, in particular cooling with non-equilibrium ionization.  A full
3D, or at least 2D calculation using realistic magnetohydrodynamics and
including the non-equilibrium ionization and associated cooling would be more
accurate than our calculations if sufficient spatial resolution could be
achieved. We aim to carry out such calculations in the future.  For such
calculations, assuming an initially uniform field, we would expect the results
to have a dependence on angle relative to the magnetic poles.  We would expect
our results presented here to most closely match those for parcels relatively
close to the magnetic equator of a remnant where the magnetic field is close
to being perpendicular to the shock propagation direction.  A more complex,
twisted or turbulent field in the ambient medium could lead to the magnetic
tension playing less of a role and to the SNR evolving in a way similar to our
results over much of its area where the magnetic pressure is the dominant
influence.

\subsubsection{Standard Run Parameters}
For our standard run we used an explosion energy of $5\times10^{50}$ ergs,
somewhat less than is usually considered for the initial explosion for SNRs,
though we have also done a run with twice that explosion energy.  Because of
energy losses to cosmic rays relatively early in the lifetime of a SNR, a
remnant with an initial explosion energy of $10^{51}$ ergs may resemble one
with a lower explosion energy that did not have cosmic ray losses. In addition
the explosion energy derived for some older remnants, such as the Cygnus Loop,
is below 10$^{51}$ ergs.  The density and magnetic field for our standard run
were $0.25$ cm$^{-3}$ and 3 $\mu$G and the ambient temperature was assumed to
be $10^4$ K.  We consider these values to be fairly representative of the warm
ISM and they match the values for steady shocks used by JTH96, which allows
for simpler comparisons to their results.  We have also done runs with higher
ambient density and lower magnetic field strength.  We discuss the effects of
these parameters on the amount of grain destruction in \S \ref{sect:parms}.

\begin{figure}[ht!]
\epsscale{0.8}
\plotone{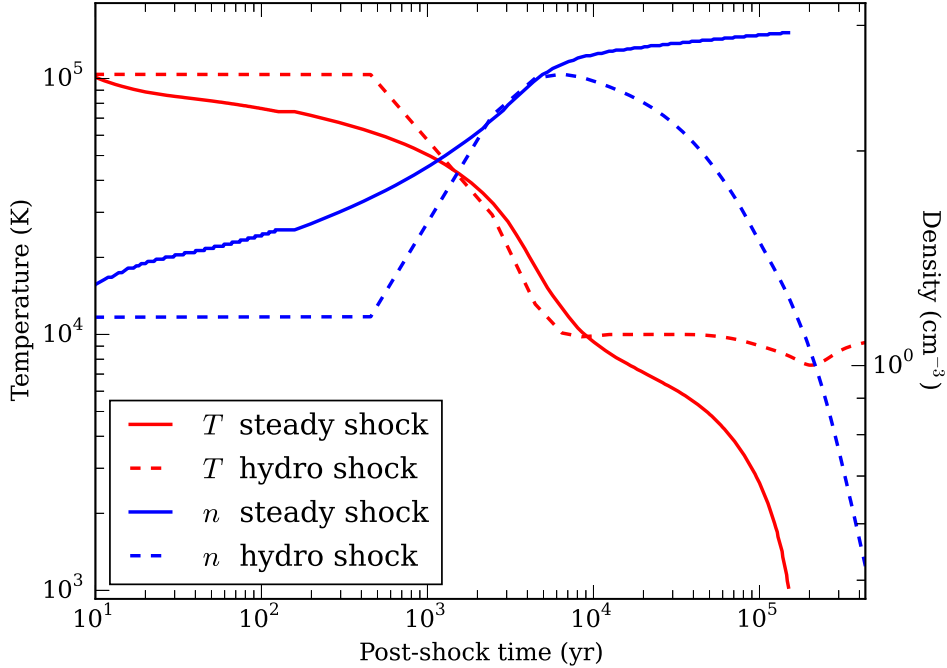}
\caption{Comparison of 100 km s$^{-1}$ shock profiles for the steady, plane
parallel shock used by JTH96 and the effective profile we calculated for
our standard hydrodynamical evolution. Note the dropoff of the density for
the hydrodynamical case.  This is caused by the spherical divergence for the
radially expanding shock.  The flattening of the hydro shock temperature
profile at $T \sim 9000$ K is caused by the switching off of cooling below
$10^4$ K.}
\label{fig:comp_shock_profs}
\end{figure}

\subsection{Parcel Trajectories}
After carrying out the hydrodynamical runs we have created parcel
``trajectories'', i.e.\ histories of the temperature, density and ionization
of individual parcels.  Using these, in addition to the shock speed as a
function of time, allowed us to create shock profiles for parcels that begin
when the shock overruns them, at a given radius.  This is made considerably
easier by the fact that the calculations are 1D, so the mass coordinate, $m(r)
\equiv \int_0^r \rho(r') 4\pi r'^2\,dr'$, for a given radius identifies the
location of a parcel.  The effective shock profiles can be compared with those
generated by steady, plane parallel shock calculations and can be used to
calculate grain destruction as the parcel moves through the shock.  In Figure
\ref{fig:comp_shock_profs} we compare the steady shock profile for a 100 km
s$^{-1}$ shock used in JTH96 with the profile we generated
for a parcel shocked to 100 km s$^{-1}$ in our standard case hydrodynamical
evolution.  The steady shock profile is done at much higher spatial resolution
than our hydro calculations, but a fairly close match is seen.  The most
striking difference between the profiles is in the density, which drops off in
the post-shock region for the hydro calculation because of the spherical
divergence.  The post-shock temperature in the hydro calculation stops
dropping and flattens out at $\sim 9000$ K because, as described above, we
turn off cooling and turn on heating for $T < 10^4$.  As we show below, the
grain destruction for these profiles turns out to be nearly the same.

\subsection{Grain Processing}
Grain processing in shocks occurs because of the acceleration of the grains
and the compression and heating of the gas. Both gas-grain interactions
(primarily sputtering) and grain-grain collisions are important in grain
destruction and erosion. The grain processing codes that we use are nearly the
same as those used by JTH96, though we have revised the dust charging scheme
as described below.  The code reads in the shock profile and integrates
through it calculating the grain processing.  The processes included are
thermal sputtering, i.e.\ sputtering caused by thermal ions, non-thermal
(a.k.a.\ inertial) sputtering caused by the relative speed of the grains and
the gas, and shattering and vaporization caused by grain-grain collisions.
We note that thermal and inertial sputtering can be combined, using a skewed
Maxwellian distribution as in \citet{Draine+Salpeter_1979a} and
\citet[hereafter BJS14]{Bocchio_etal_2014}, though we have not done that in
this work.  The differences between the two approaches is most pronounced for
case in which the thermal velocity of the gas and drift velocities of the
grains are similar.  The sputtering rates we use are the same as in JTH96.
BJS14 have used updated rates for carbonaceous grains based on the work of
\citet{SerraDiaz-Cano+Jones_2008}, with the assumption that the carbonaceous
grains are hydrogenated amorphous carbon (a-C:H).  We use the older rates that
apply to graphitic grains in this work.  The assumption that the carbonaceous
grains are a-C:H type leads to more destruction as we discuss below. 

Because shattering re-distributes grain mass into smaller grains, mass bins
must be followed (see JTH96 for details). For most of the calculations, we
divide the grain mass into five bins initially, with an extra bin on the low
end to contain small grains that result from shattering or sputtering. JTH96
found that there was little difference in the grain destruction results if a
larger number of bins were used. We have found this to be the case for our
calculations as well.  However, to make the visualization of the redistribution
of grain mass caused by shattering clearer, we have also carried out
calculations using an initial seventeen mass bins with seven initially empty
small mass bins for our standard run. 

\subsubsection{Grain Parameters}
The grain parameters that we adopt are the same as those used by JTH96 so as
to facilitate comparison with their results.  We use two grain types,
silicates and carbonaceous grains with assumed solid material densities of 3.3
and 2.2 g cm$^{-3}$ respectively. We assume an initial \citet[hereafter
MRN]{Mathis_etal_1977} type power law size distribution ($dn/da \propto
a^{-3.5}$) for both the carbonaceous and silicate grains. The smallest and
largest grain sizes at the start of the grain evolution calculations are 0.005
and 0.25 $\mu$m for both grain types used.  The assumed gas-to-dust mass ratio
in the ambient medium (including He in the gas mass) is 163. The included dust
destruction processes, i.e.\ thermal sputtering, non-thermal sputtering,
shattering and vaporization, involve a number of physical parameters and we
refer the reader to JTH96 for a detailed discussion of them.  Here we just
note that the assumed atomic mass for silicate grains is 5.7 u and 4 u for
the carbonaceous grains and the threshold velocity for shattering is 2.7
km s$^{-1}$ and 1.2 km s$^{-1}$ for silicate and carbonaceous grains
respectively.  Much of the destruction in radiative shocks begins with
shattering, which creates small grains which are then sputtered, though also
slowed more quickly than larger grains.

We acknowledge that this model for the size distribution and
grain composition is not consistent with all of the observational constraints.
More recent models \citep{Weingartner+Draine_2001,Zubko_etal_2004,
Jones_etal_2013,Siebenmorgan_etal_2014} have been constructed that are better
able to fit the various observational constraints on the dust such as the
infrared emission and abundance constraints as well as the extinction curve.
Those models use more complicated grain size distributions and more realistic
grain compositions.  In the future we aim to test the effects of these grain
models on grain destruction timescales, but in this work our primary aim has
been to compare with the steady state plane parallel shock calculations that
have been carried out previously and therefore we have used the same grain
model as in JTH96.  

\subsubsection{Grain Charge}
The interaction of dust with the plasma depends in various ways on its charge.
JTH96 used the rough analytical fit of \citet{McKee_etal_1987} to the results
of charging calculations to determine the dependence of grain charge on the
gas temperature, relative gas-grain speed and UV radiation field.  We have
updated the grain charge calculations using the scheme from
\citet{Weingartner+Draine_2001} with updates from
\citet{Weingartner_etal_2006} and modifications to include the relative
gas-grain velocity, which alters the charging rates for electron and ion grain
collisions.  In this work we adopt the standard tight coupling approximation,
in which it is assumed that the gas and grains remain coupled throughout the
shock. Under these conditions, the value of the grain charge has only a
limited role.  The grain undergoes betatron acceleration when the magnetic
field increases because of compression, but though this is mediated by the
grain charge, the actual value of the charge does not come into the
calculation of the grain speed.  The only effect of the charge in this scheme
is in its effect on the plasma drag on the grains and the small Coulombic
modification to the cross section for sputtering.  We note that the grain
charge can be much more important if the gas and dust decouple as can occur
for larger grains \citep[see][]{Slavin_etal_2004}.

An input that is needed for the grain charging calculations is the
yield-averaged UV flux.  Following \citet{McKee_etal_1987} we define
\begin{equation}
G_0 = \frac{\int_{8 \mathrm{eV}}^{\infty} d\nu \,Y(h\nu)\, 4\pi J_\nu/h\nu}{2.4\times10^6 \;\mathrm{cm}^{-2}\, \mathrm{s}^{-1}},
\end{equation}
where $Y$ is the photoelectric yield and $J_\nu$ is the mean intensity of the
radiation field and $2.4\times10^6$ cm$^{-2}$ s$^{-1}$ is the ``average''
yield averaged UV field as adopted by \citet{Draine+Salpeter_1979b}. In
principle, given the temperature, density and ionization at every point in the
remnant as a function of time, one could calculate this quantity at each
radius and time.  However, in detail it becomes more complex since the largest
contributors to $J_\nu$ are resonance lines, in particular H Lyman
$\alpha$, that have substantial optical depth in the shock region.  Thus
scattering is important and an accurate calculation would require a radiative
transfer calculation for each location and time.  Given the limited role of
grain charge for the grain destruction calculations, we take a simpler
approach.  Following \citet{McKee_etal_1987} we take $4\pi J(\mathrm{Ly}
\alpha) = 30 F(\mathrm{Ly} \alpha)$, where $F(\mathrm{Ly} \alpha)$, is one
half the integrated column emissivity of H Lyman $\alpha$ in the shock and
$J(\mathrm{Ly} \alpha)$ is the mean intensity.  The factor of 30 here comes
from both geometrical effects (effective extent of the front perpendicular to
the shock, a factor of $\sim 3$) and enhancement caused by scattering (a
factor of $\sim 10$). In addition we assume that there is a diffuse background
equal to the \citet{Draine+Salpeter_1979b} field.  Doing this we find for our
standard run that $G_0$ has a peak of $\sim 2.7$ for shock speeds near 115 km
s$^{-1}$ falling off sharply at higher velocities and more gradually toward
lower speeds, reaching 1.4 at $v_\mathrm{shock} = 40$ km s$^{-1}$, which is
the shock speed at the end of the standard run.  These values are somewhat
higher than those reported by \citet{McKee_etal_1987}, though that could be
caused, especially for the slower shocks, by the effects of earlier shocks
that affect the post-shock gas that was shocked to a higher speed.  
The effects of the radiation field on the grain destruction is minor, since in
tests we have done, an increase by a factor of ten on $G_0$ did not change the
overall grain destruction efficiency substantially.  

\section{Results}
\subsection{SNR Evolution}
To assess the impact of SNR shocks on dust destruction one needs to follow SNR
evolution well beyond the point at which the remnant becomes radiative.  This
is because not only do the relatively slow shocks of late stage remnants still
destroy dust, they also sweep up increasing volumes of the ISM such that even
with a low efficiency of grain destruction, slow shocks are important for the
overall grain destruction in the ISM.  In addition, for our calculations we
need to follow the gas parcels that have been shocked for a long enough time
that the grain destruction rate has become negligible.  For our standard run
we follow the evolution until the shock speed drops to 41 km s$^{-1}$
at a radius of 62 pc, $5.4\times10^5$ yr after the explosion.  The
magnetosonic speed in the ambient medium is 17.2 km s$^{-1}$, so it is a mach
2.4 shock with a post-shock maximum compression of 3.9 (after cooling).  We
note that this compression factor is less than estimated by
\citet{McKee_etal_1987} for a magnetically supported post-shock, compression
region, $\chi_\mathrm{max} = 0.767 v_\mathrm{shock} n_a^{1/2}/B_a$, where
$\chi$ is the compression factor, $v_\mathrm{shock}$ is the shock speed in km
s$^{-1}$, $n_a$ is the ambient density in cm$^{-3}$ and $B_a$ is the ambient
magnetic field in $\mu$G.  That expression yields a value of 5.24 for this
shock. We find that in general the maximum post-shock compression in our
hydrodynamical calculations is below that predicted by McKee et al.'s
expression since thermal pressure also contributes to the post-shock pressure
and the spherical divergence and expansion over time causes the pressure to
decline behind the shock.

\begin{figure}[ht!]
\epsscale{0.8}
\plotone{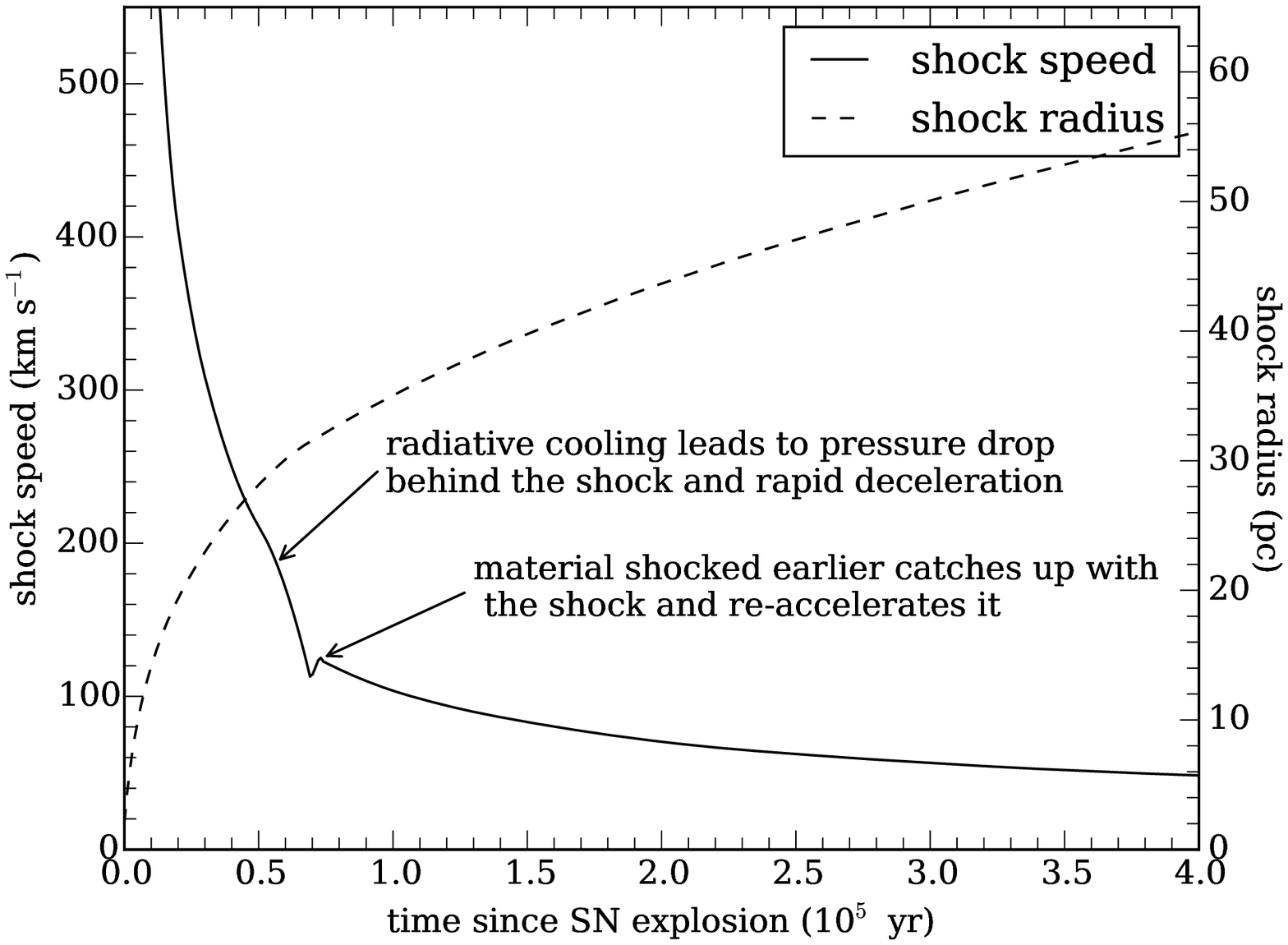}
\caption{Shock radius and shock velocity evolution for our standard, supernova
remnant run ($E_0 = 0.5\times 10^{51}$ ergs, $n_a = 0.25$ cm$^{-3}$, $B_a =
3\,\mu$G). Once the remnant begins to go radiative at $t \approx 5\times 10^4$
yr, the shock speed drops sharply and then rebounds after previously shocked
material catches up with the decelerating shock.  As a result of the sharp
drop in shock speed there is a range of speeds, $v_\mathrm{shock} \approx 120
- 190$ km s$^{-1}$, that exists for only a short time and in a small volume of
the ISM.}
\label{fig:vs_Rs}
\end{figure}

\begin{figure}[ht!]
\epsscale{1.}
\plotone{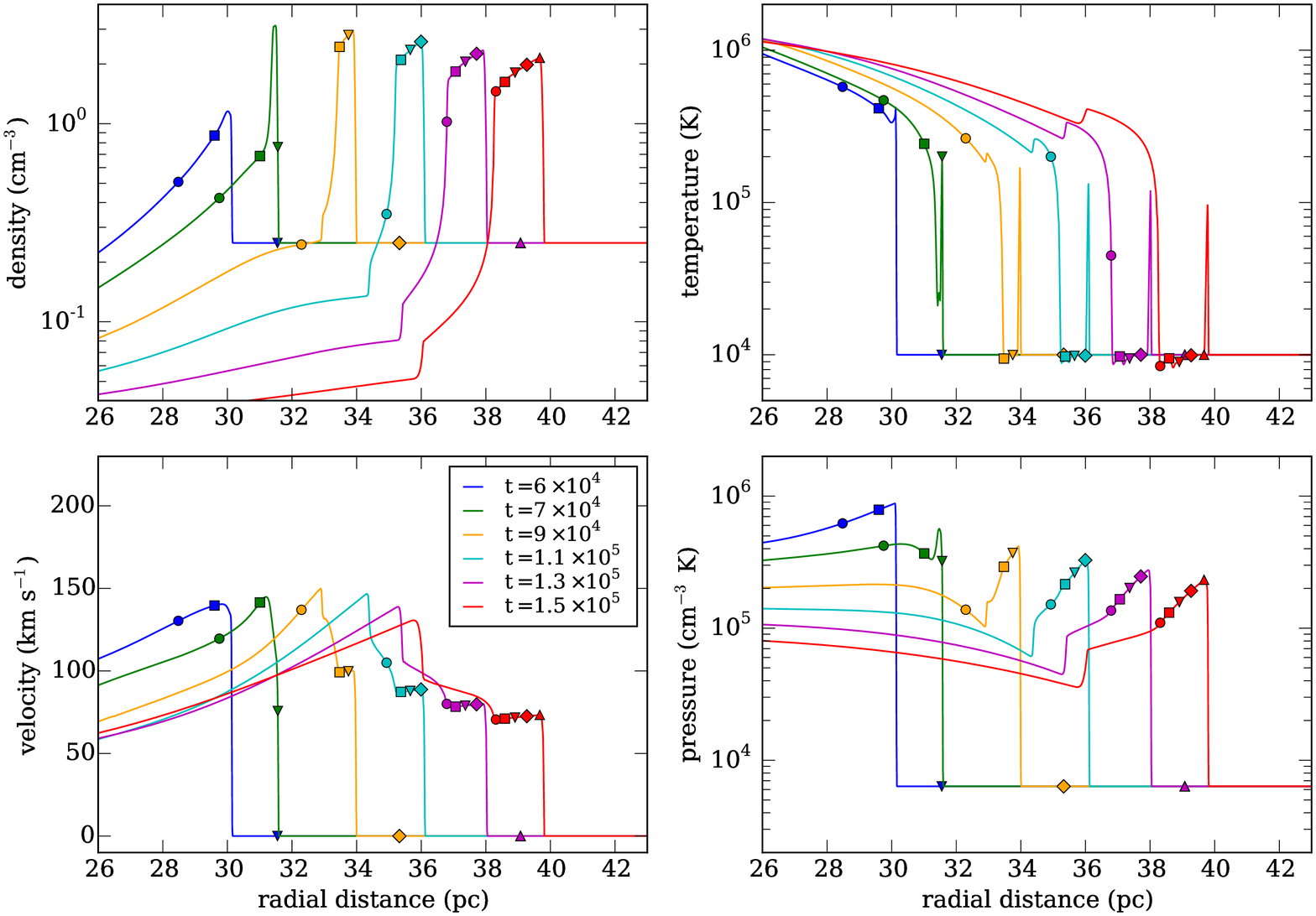}
\caption{Evolution of individual gas parcels within our standard SNR evolution
run.  Gas parcels are identified by their marker shape, while the colors
correspond to time snapshots.  The initial radii of the parcels are 24.075,
27.825, 31.550, 35.325 and 39.075 pc from the center of the SNR.  The
corresponding speeds to which these parcels were initially shocked are:
285, 215, 114, 102, and 86 km s$^{-1}$, respectively.  See text for more
discussion.  Note that the pressures shown in the lower right panel are the
total pressures including thermal and magnetic pressures.}
\label{fig:parcel_evol}
\end{figure}

Figure \ref{fig:vs_Rs} shows the shock radius and shock speed evolution over
time for our standard run.  The precipitous drop in shock speed at just below
200 km s$^{-1}$ is clearly visible and indicated on the plot.  As shown the
shock is briefly re-accelerated when the previously shocked material catches
up with the slowing cold shell.  This leads to greater compression at this
time and cooling of the hot shocked gas just inside of the cooled radiative
shell as it runs into and is compressed against the shell.  This is
illustrated in Figure \ref{fig:parcel_evol} which shows the evolution of
gas parcels.

\begin{figure}[ht!]
\epsscale{1.}
\plotone{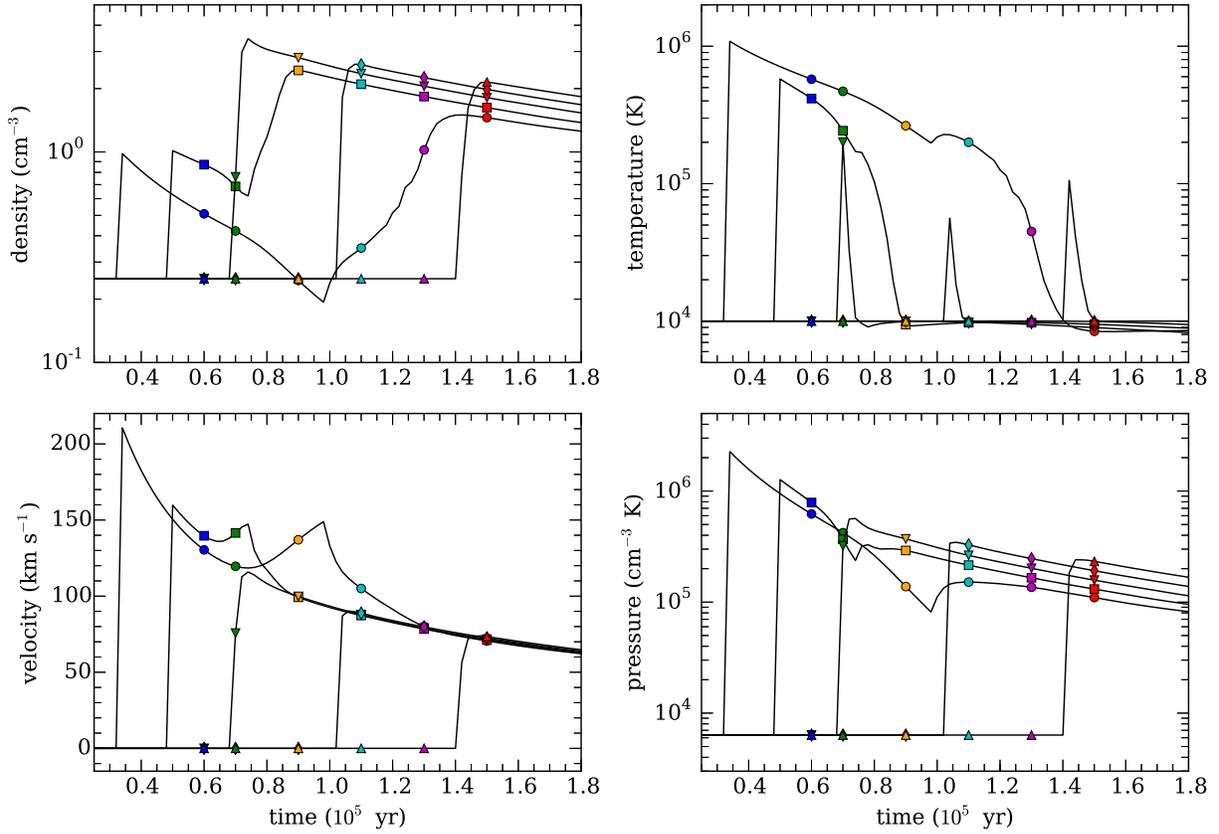}
\caption{Evolution of individual gas parcels with time.  Here we show the time
evolution of the fluid variables for the same parcels as shown in
Figure \ref{fig:parcel_evol} with the same marker symbols.  The color of the
symbol corresponds with the time as in Figure \ref{fig:parcel_evol}.  These
are the shock profiles that are input into the grain evolution code that
calculates grain processing including destruction via sputtering and
vaporization and shattering (though the time used for each profile is relative
to when the parcel is first shocked).  Note in particular the non-monotonic
evolution of the density and temperature for the parcel represented by a
circle.  The blastwave velocity was 285 km s$^{-1}$ when this parcel was
shocked and it was heated to $1.1\times 10^6$ K.  Even many parcels that were
shocked to high temperatures initially, cool well before their initial
post-shock radiative cooling times would predict because of being compressed
against the dense shell.}
\label{fig:parcel_traj}
\end{figure}

Figure \ref{fig:parcel_evol} and Figure \ref{fig:parcel_traj} illustrate how gas
parcels shocked to a range of shock speeds evolve within the remnant.  The
parcels highlighted in the figures are at initial radii ranging from 24.075 pc
to 39.075 pc and are shocked to speeds ranging from 86 to 285 km s$^{-1}$.
The first time shown, $6\times10^4$ yr after the explosion, is at a point when
the remnant is just beginning to go radiative, as can be seen in Figure
\ref{fig:parcel_evol} from the density profile, which has a peak value a
little greater than 4 times the ambient density of 0.25 cm$^{-3}$.  By
following the different parcels over time one can see that the density and
temperature evolution is complicated for some parcels.  This can be seen
clearly in Figure \ref{fig:parcel_traj} which shows the time evolution of the
same parcels.  In particular, the parcel marked by circles (first parcel to be
shocked) can be seen to be compressed and heated, then expand (density
decreases) and finally get recompressed and cooled because of catching up with
the cold shell.

There have been many numerical calculations of SNR evolution carried out over
the years
\citep[e.g.,][]{Cox_1972,Chevalier_1974,Cioffi_etal_1988,Slavin+Cox_1992}.
\citet{Cioffi_etal_1988} analyzed the results of their 1D (spherically
symmetric) numerical calculation of radiative remnant evolution, which did not
include any magnetic pressure, and found that the late time shock radius
evolution could be well described by an offset power-law, $R_s =
R_\mathrm{PDS} (4/3 t_* - 1/3)^{3/10}$,  where $R_\mathrm{PDS}$ and $t_*$ are
functions of the explosion energy, ambient density and metallicity.  We find a
that we can also fit the late time shock radius evolution with such an offset
power law, but with a somewhat different exponent.  For our standard case we
find the late time behavior is not well fit by the \citet{Cioffi_etal_1988}
formula and instead fit a general offset power law, $R_s \propto (t -
t_\mathrm{off})^\gamma$ to the expansion.  This yields $R_s = 0.494 (t +
1.79\times 10^4 \mathrm{yr})^{0.365}$ pc.  Note in particular that we get an
offset time with the opposite sign from that in the Cioffi et al.\ formula.
In general we expect that including magnetic pressure will lead to faster
expansion since the total pressure in the shell is dominated by magnetic
pressure.  The late time expansion is thus maintained primarily by the energy
stored in the magnetic field of the shell rather than the thermal energy in
the hot bubble.  Since the overall grain destruction rate depends on the mass
of gas shocked to a given speed, $M(v_s)$, the shock speed as a function of
radial distance from the explosion center is important.  We discuss this
further below.

\begin{deluxetable}{crrrrrrr}
\tablecolumns{8}
\tablewidth{0pc}
\tablecaption{Comparison of Modeled Grain Destruction Efficiencies (\%)}
\tablehead{ & \multicolumn{7}{c}{$v_\mathrm{shock}$ (km s$^{-1}$)}\\
\cline{2-8}
\colhead{Ref.\tablenotemark{a}} & \colhead{50} & \colhead{75} &\colhead{100} &
\colhead{125} & \colhead{150} & \colhead{175} & \colhead{200}}
\startdata
\sidehead{Carbonaceous grains}
\tableline
JTH96      &  1 &  5 &  7 & 13 & 12 &  21 &  47 \\
BJS14      & 77 & 83 & 91 & 96 & 99 & 100 & 100 \\
this study &  1 &  4 & 10 & 18 & 17 &  18 &  23 \\
\tableline
\sidehead{Silicate grains}
\tableline
JTH96      & 2 & 12 & 18 & 33 & 32 & 41 & 49 \\
BJS14      & 2 & 12 & 29 & 46 & 53 & 67 & 67 \\
this study & 2 &  9 & 23 & 40 & 41 & 42 & 40 \\
\enddata
\tablenotetext{a}{The references correspond to: JTH96, \citet{Jones_etal_1996},
and BJS14, \citet{Bocchio_etal_2014}. ``this study'' refers to our standard
run that uses the parameters in the first line of Table \ref{tab:mass_dest}.
Note that the results for the 75, 125 and 175 km s$^{-1}$ shock speeds were
not presented in JTH96 but were calculated using the same code later.}
\label{tab:comp_eff}
\end{deluxetable}

\begin{figure}
\plotone{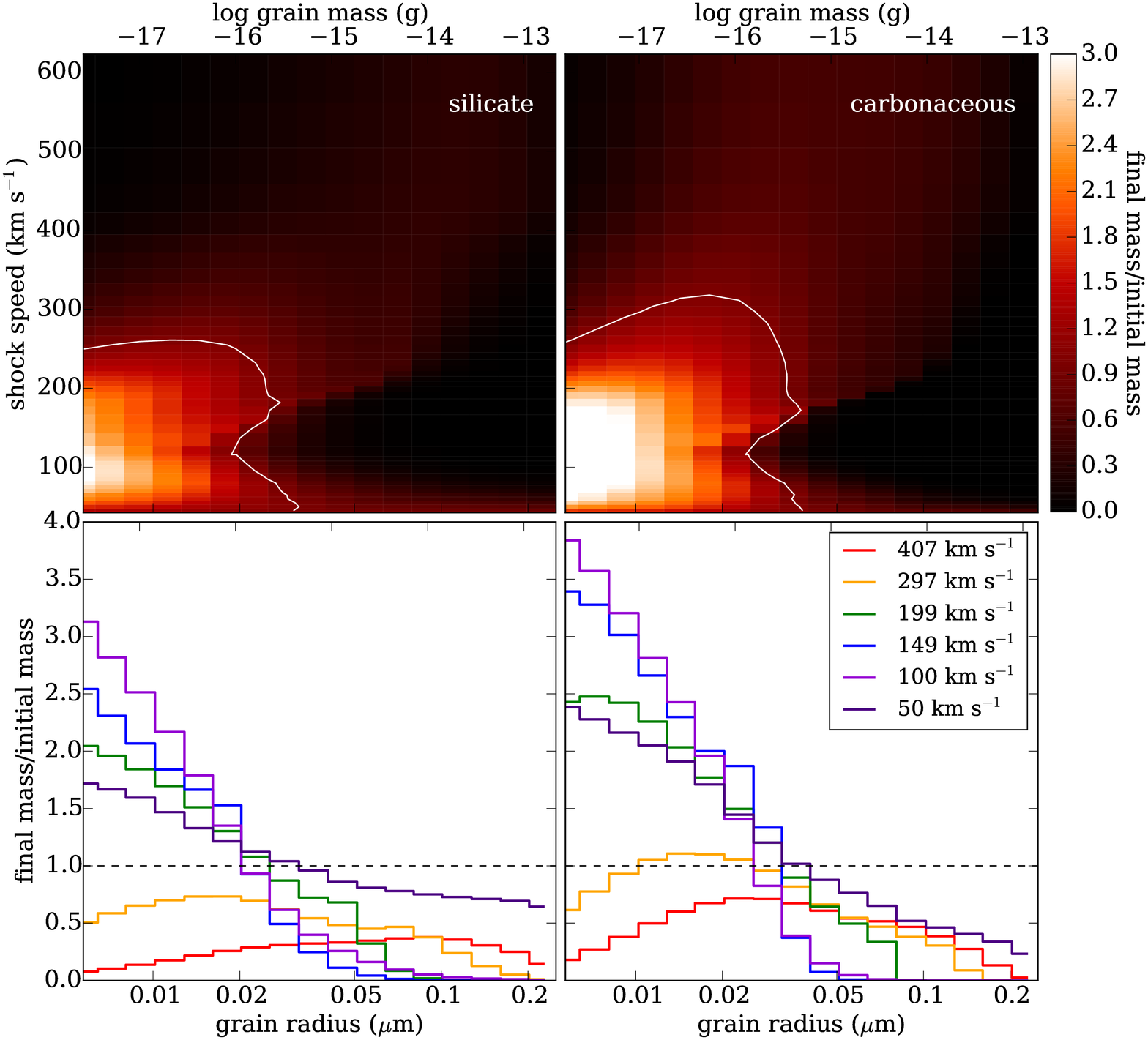}
\caption{Ratio of final grain mass/initial grain mass in size/mass bins as
a function of shock speed and grain size (top) for silicate grains (left) and
carbonaceous grains (right).  The bottom plots show the same data as line
plots of mass ratio as a function of size/mass bin for selected shock
speeds.  The white contours in the top plots show the division between
bins for which there is a net gain in mass and those for which there is a net
loss.  The dashed line in the bottom plot shows the same division between net
increase and decrease of mass in a size bin.  The size bins are
logarithmically spaced in mass.  Although it may appear that there is an
overall increase in mass for some shocks, the increase in each bin is relative
to the initial value and since most of the mass is in the large size/high mass
bins initially, so in fact dust mass is lost in all cases.  These results are
for our standard run.}
\label{fig:fracdest_imgs}
\end{figure}

\begin{figure}
\epsscale{1.1}
\plottwo{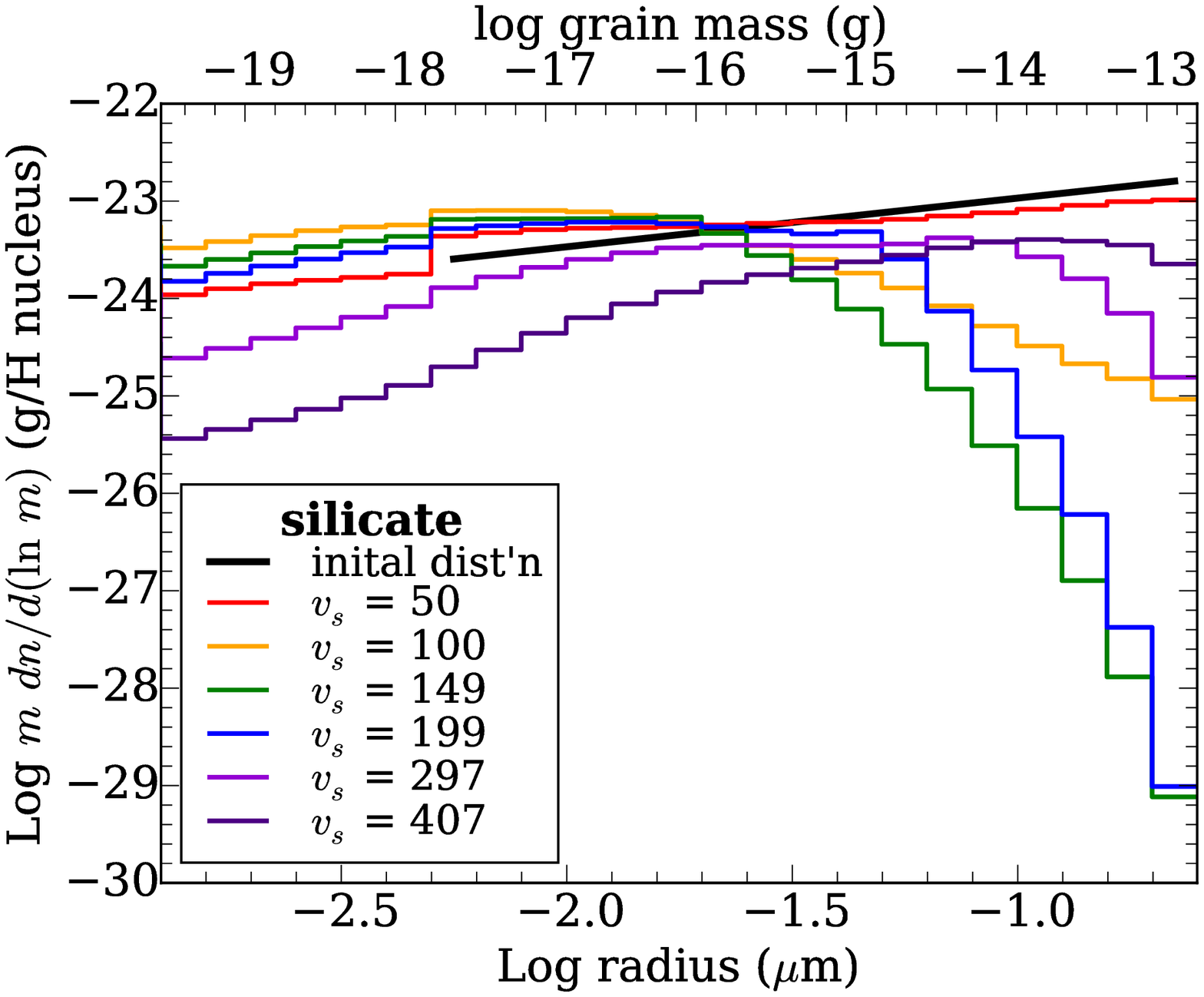}{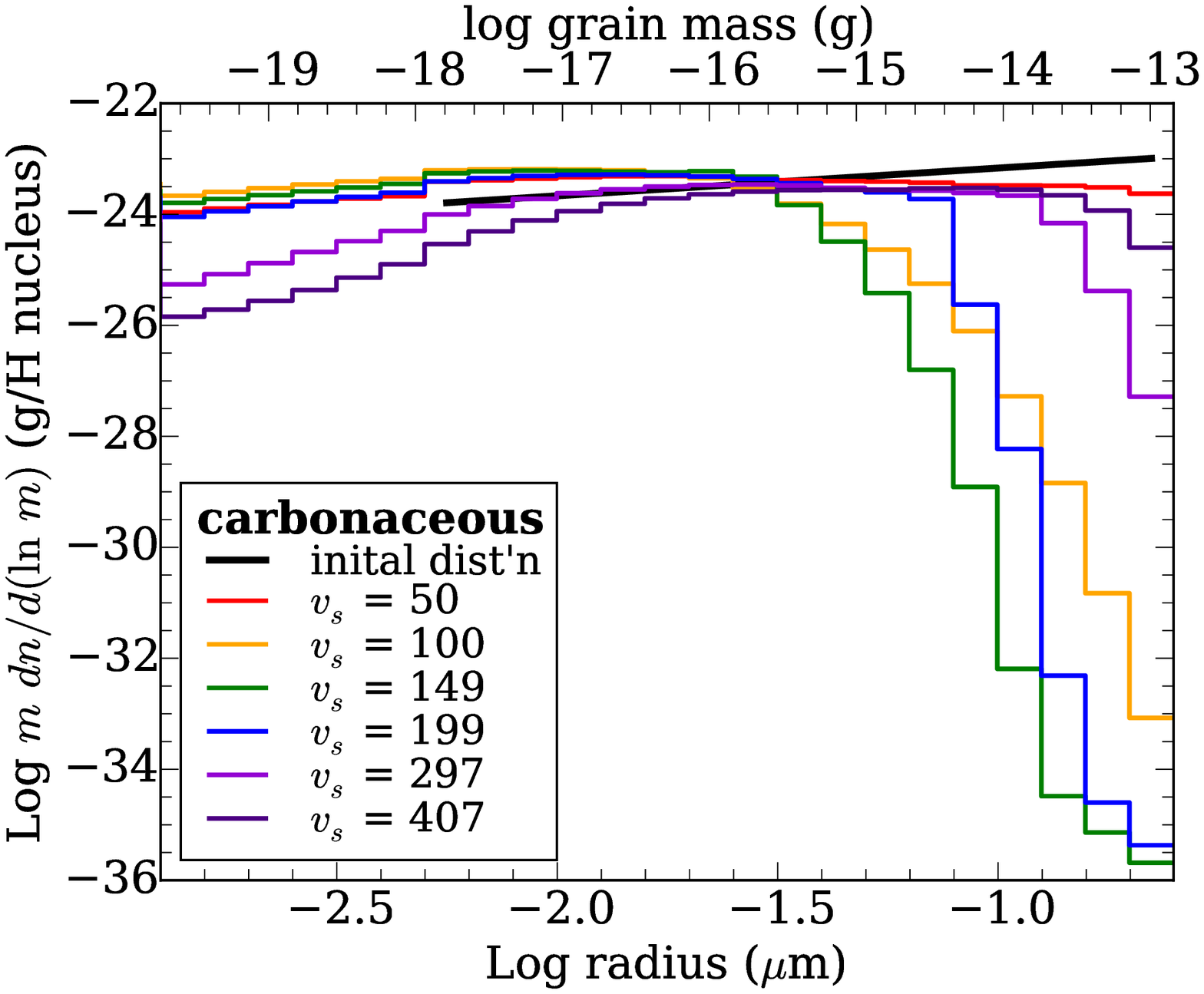}
\caption{Grain mass distribution vs.\ grain size and mass for several
different shock speeds for our standard hydro run.  This version of the size
distribution illustrates that most of the mass is at the high mass end of the
size distribution and is proportional to $a^4\,dn/da$, where $dn/da$ is the
standard size distribution.  Note that the initial MRN-type size distribution
extends only from 0.005 $\mu$m to 0.25 $\mu$m, but after shock processing much
of the grain mass gets shifted to smaller (lower mass) grains primarily
because of shattering.}
\label{fig:mass_distn}
\end{figure}

\subsection{Grain Destruction Efficiency}
The grain destruction efficiency is defined as the fraction of the mass
originally contained in grains that is returned to the gas phase either
through sputtering or vaporization of the grains.  Grains that are fragmented
are not considered to be destroyed.  Their mass is distributed into lower mass
bins (see JTH96 for a more thorough discussion and Figure
\ref{fig:mass_distn}).  In Figure \ref{fig:fracdest_imgs} and
\ref{fig:mass_distn} we illustrate the destruction and re-distribution of
grain mass for shocks of a range of speeds for our standard run.
In Figure \ref{fig:fracdest_imgs} we plot, instead of the destruction
efficiency, the ratio of final mass over initial mass in each bin,
$m_{kf}/m_{ki}$ where $m_{ki}$ is the initial mass contained in bin $k$ and
$m_{kf}$ is the final mass contained in the bin.  Mass is lost to a bin in one
of three ways: 1) via sputtering (in which case the sputtered mass fraction is
lost to the dust phase), which causes grains to move down to lower mass bins,
2) vaporization, which causes the entire grain mass to be lost to the dust
phase or 3) shattering, in which case no mass is lost to the dust phase, but
the mass is redistributed to the lower mass bins.  Altogether these processes
tend to cause a substantial loss of large grain mass, mostly from shattering,
with the mass in small grains increasing relative to the initial size
distribution.  These results are illustrated in Figure \ref{fig:fracdest_imgs}
where, in the upper plots, the white line divides the area in which mass is
removed from the mass bins from that in which there is a net gain of mass.
Because of the non-monotonic shock speed evolution, we cannot show the results
on these plots for all the shocks with speeds $\sim 125$ km s$^{-1}$ and we
simply choose the first time a shock speed was encountered for display
purposes.  Since in the plots in Figure \ref{fig:fracdest_imgs} we are
plotting the ratio relative to the initial mass, we cannot include bins that
are initially empty.  For that reason we also plot the absolute mass
distributions in Figure \ref{fig:mass_distn}.  More specifically we plot
$m\,dn/d(\ln m) = m a/3\, dn/da$, where $m$ is the grain mass and $dn/da$ is
the size distribution initially $\propto a^{-3.5}$ (MRN distribution).  

\begin{figure}[ht!]
\epsscale{0.8}
\plotone{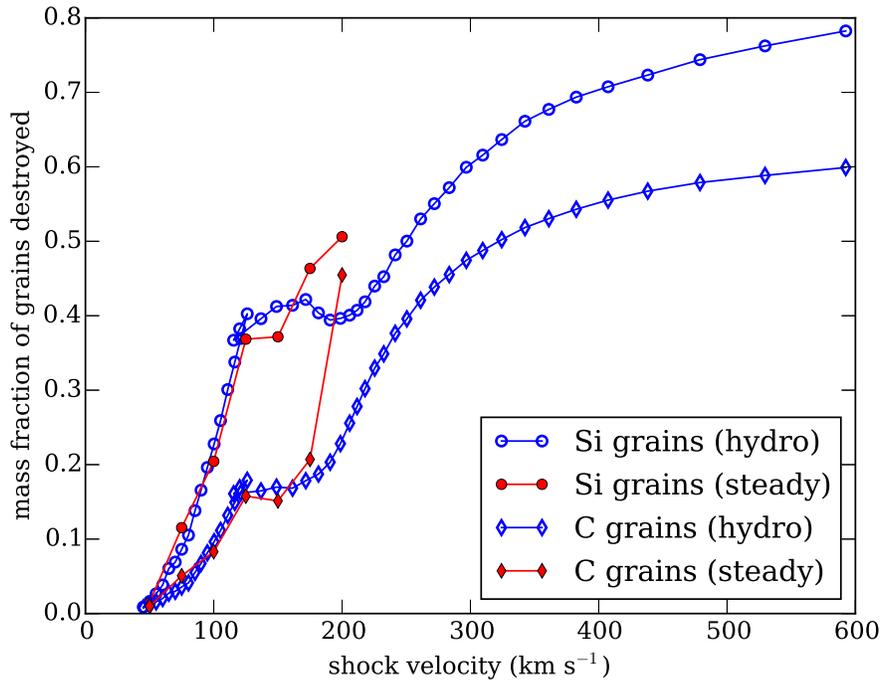}
\caption{Grain destruction efficiency vs.\ shock velocity for our standard
hydrodynamical run compared to that for steady, plane parallel shocks. The
results for our hydro runs are clearly quite close to those for the steady
shocks except for near $v_s \approx 200$ km s$^{-1}$ where the hydro runs
yield a substantially smaller efficiency.  The cause of this lower efficiency
is the early onset of compression because of running into the rapidly slowing
shock in the hydrodynamical evolution.  The magnetic field causes a partial
re-bound before cooling sets in followed again by expansion because of
spherical divergence.  These effects lead to less compression and therefore
less grain acceleration and destruction in the shock.}
\label{fig:eps_vshock}  
\end{figure}

We show in Figure \ref{fig:eps_vshock} our calculated destruction efficiency
vs.\ shock speed with a comparison to the results for steady plane-parallel
shocks as in JTH96.  In Table \ref{tab:comp_eff} we compare our results with
those of JTH96 and BJS14. The large differences between the efficiency of
carbonaceous grain destruction found by BJS14 and both JTH96 and this study
derive from the different grain model used by that work.  Rather than the
graphite-like carbonaceous grains used by JTH96, BJS14 use hydrogenated
amorphous carbon (a-C:H) as their model for cabonaceous grains and employ the
sputtering yield results of \citet{SerraDiaz-Cano+Jones_2008}.  The sputtering
yield for this type of carbonaceous grain is substantially greater than that
for graphite due to a lower binding energy and density for a-C:H.  In
addition, BJS14 use different size distributions for both the carbonaceous and
silicate grains.  Since grain-grain collisions are important for altering the
grain size distribution in shocks, a different initial size distribution can
have a substantial impact on the grain processing.  Finally, BJS14 include
mantles on the silicate grains entering the shock.  These are features
of the \citet{Jones_etal_2013} dust model, which differs significantly from
the MRN model used by JTH96 and by us in this work.

We have recalculated the steady shock efficiencies using a slightly modified
version of the GRASH code employed by JTH96, using updated grain charge
calculations as discussed above.  We have also supplemented the original four
steady shock speeds (50, 100, 150 and 200 km s$^{-1}$) with three additional
shock speeds (75, 125, and 175 km s$^{-1}$) using shock profiles kindly
provided to us by John Raymond.  As we have noted, for an evolving SNR that
goes radiative, shock speed does not evolve monotonically because of the
sudden drop in post-shock pressure and shock speed and subsequent
re-acceleration that occurs at $v_s \sim 125$ km s$^{-1}$.  This is why the
efficiency vs.\ $v_s$ curves are multivalued in the figure.  As can be seen in
the figure, the hydrodynamical calculations agree with the steady, plane
parallel shock results quite closely for most of the range covered by the
steady shock calculations.  The exception is at the high shock speed end.  The
difference for these shocks seems to be caused by the non-steady hydro
evolution when the radiative losses are just starting to become important.
For this period, the shocked gas does not radiate immediately but cools slowly
at first, only forming a cold shell after $\sim 10^4$ yr.  In the hydro case,
compression is initiated by the shocked gas catching up to the slowing shell,
because later, slower shocks go radiative much more quickly.  Thus for the
shocks with speeds near 200 km s$^{-1}$, the gas is first compressed and only
then cools because of the higher density.  The compression causes an increase
in magnetic pressure that causes a back reaction that causes the shell to
re-expand somewhat before it has fully cooled.  This prevents the post-shock
gas from reaching the high densities that it does for steady shock evolution
and therefore the grains do not get as much acceleration with accompanying
destruction.

\begin{deluxetable}{ccccrcc}
\tablecolumns{7}
\tablewidth{0pc}
\tablecaption{Mass of Gas with Cleared of Dust for SNR Models\tablenotemark{a}}
\tablehead{ \colhead{$E_{51}$} & \colhead{$n_{\mathrm{a}}$ (cm$^{-3}$)}
& \colhead{$B_{\mathrm{a}}\,(\mu$G)} &
\colhead{$X_\mathrm{H}$\tablenotemark{b}} & \colhead{TC\tablenotemark{c}} & 
\colhead{$M_\mathrm{Si}$ (M$_\odot$)} & \colhead{$M_\mathrm{C}$ (M$_\odot$)}}
\startdata
0.5 & 0.25 & 3.0 & 0.5 & on & \phn990 & \phn600 \\
0.5 & 0.25 & 3.0 & 0.1 & on & \phn980 & \phn600 \\
0.5 & 0.25 & 3.0 & 0.5 & off & \phn960 & \phn580 \\
1.0 & 0.25 & 3.0 & 0.5 & on & 1990 & 1220 \\
0.5 & 0.25 & 0.3 & 0.5 & on & 1490 & \phn780 \\
0.5 & 1.00 & 3.0 & 0.5 & on & 1020 & \phn640 \\
\tableline
\sidehead{Steady shock/HIM dominated\tablenotemark{d}}
\tableline
1.0 & 0.25 & 3.0 & \nodata & \nodata & 1700 & \phn990 \\
\enddata
\tablenotetext{a}{The masses are the efficiency of dust destruction (fraction
of grain mass destroyed) integrated over gas mass enclosed by the shock over
the calculated SNR evolution (see Eq.\ \ref{eqn:mdest} and
text). The first line in the table corresponds to our standard case.}
\tablenotetext{b}{Hydrogen ionization fraction in the ambient medium.  Helium
is assumed to have the same ionization fraction.}
\tablenotetext{c}{Thermal conduction --- either turned on or off.}
\tablenotetext{d}{These values are for the hot gas dominated ISM model as
discussed in JTH96 and in the text (\S\S \ref{sect:ISM_mod} and
\ref{sect:fillfact}) and include a factor of $f_c/f_h = 0.3/0.7$ (the
ratio of warm cloud to hot gas filling factors; values without this factor
are 4000 and 2300 M$_\odot$ for silicate and carbonaceous grains
respectively).  We used a somewhat different interpolation/extrapolation of
the grain destruction efficiency than JTH96 to derive these results (see
text).}
\label{tab:mass_dest}
\end{deluxetable}

To understand the efficiency of dust destruction in the ISM we must use the
efficiency for the evolving SNR shocks in conjunction with information on the
expansion of the SNR.  The mass of gas in which the dust has been destroyed can
be calculated by \citep[e.g.,][]{Dwek+Scalo_1980}
\begin{equation}
M_g = \int \epsilon(M_s) \, dM_s,
\label{eqn:mdest}
\end{equation}
where $\epsilon$ is the grain destruction efficiency and $M_s$ is the mass of
shocked gas.  This has been sometimes been converted to an integral over
$v_s$, $\int \epsilon(v_s)\, dM_s/dv_s\; dv_s$, but this cannot be done easily
when $v_s$ is not monotonic with $M_s$ as is our case.  This integral gets cut
off at the low (mass) end by the small amount of gas that is shocked to high
velocities and at the high end by the low grain destruction efficiency at low
shock velocities, which goes to zero as $v_s$ approaches the signal speed of
the ambient medium.  We list the mass of gas cleared of dust in Table
\ref{tab:mass_dest}.  We did not find a simple fitting formula for $\epsilon$
for our standard hydro run, but it can be approximated by
\begin{equation}
\epsilon(v_{s7}) = \left\{ \begin{array}{ll}
0.05 - 0.19\, v_{s7} + 0.215\, v_{s7}^2 + 0.0200\, v_{s7}^3 & 0.4 < v_{s7} < 1.25 \\
0.175 & 1.25 \le v_{s7} < 1.85 \\
-1.9 + 2.02\, v_{s7} - 0.641\, v_{s7}^2 + 0.0922\, v_{s7}^3 - 0.00498\,
v_{s7}^4 & 1.85 \le v_{s7} < 5 
\end{array}
\right.
\end{equation}
where $v_{s7} = v_s/100$ km s$^{-1}$. 

In addition to our results for our hydrodynamical calculations, we present in
Table \ref{tab:mass_dest} results for steady shock calculations as in JTH96.
For those models, the assumed pre-shock ionization depended on the shock speed
and so is not listed in the table.  JTH96 provided an approximate fit to their
destruction efficiency results that included an extrapolation to 300 km
s$^{-1}$. The fit was based on results only for 50, 100, 150 and 200 km
s$^{-1}$ shocks.  We have used our calculated steady shock values at 75, 125,
and 175 km s$^{-1}$ to get a more accurate interpolation and have used an
almost flat extrapolation to 500 km s$^{-1}$ rather than their sharply rising
extrapolation.  The newly calculated interpolation and values tend to lie
above the old fit, especially near 125 km s$^{-1}$, which contributes
significantly to the integral for $M_g$.  Our extrapolation for shock speeds
above 200 km s$^{-1}$ is probably a lower limit, though at those high shock
speeds the contribution to the integral is not very large.  As noted in
footnote d of the table we include the factor of $f_c/f_h =
0.3/0.7$, the ratio of warm cloud to hot gas filling factors, in the mass
calculated for the steady shocks. The reasoning behind this and the importance
of the ISM model for grain destruction estimates is discussed in \S
\ref{sect:ISM_mod}.  Also mentioned in the footnote is that the values without
this factor are roughly 4000 and 2300 M$_\odot$ for silicate and carbon
grains respectively.  These values are roughly 10 - 15\% higher than the JTH96
results.  

\subsection{Effects of Parameter Choices\label{sect:parms}}
In order to assess the effects of the various assumptions that we have made
for our standard run, we have done a series of hydrodynamical runs with
different parameters.  Besides our standard run, which corresponds to the
first line in Table \ref{tab:mass_dest}, we tested the effects of changing the
pre-ionization in the ambient medium ($X_\mathrm{H}$), turning off thermal
conduction in the hot gas, higher explosion energy ($E_{51} = 1$), lower
magnetic field ($B_\mathrm{a} = 0.3 \,\mu$G), and higher ambient density
($n_\mathrm{a} = 1.0$ cm$^{-3}$).  We find that neither changing the
pre-ionization nor turning off thermal conduction have significant effects on
the destroyed dust mass.  Doubling the explosion energy approximately doubles
the amount of dust mass destroyed.  This is to be expected since the mass
shocked to a given velocity is directly proportional to the explosion energy
in the adiabatic (Sedov-Taylor) case. For radiative remnants this scaling is
not guaranteed, but \citet{Cioffi_etal_1988} find this scaling to be
approximately true for their radiative remnant calculations and we also find
that that scaling is roughly obeyed. There is also a strong effect of
decreasing the magnetic field.  Since the shell is supported by the field
after the shock goes radiative, the amount compression goes roughly inversely
with $B_\mathrm{a}$.  Thus the lower field leads to higher compression, more
grain acceleration and more grain destruction.  In our results, the increase
is about 50\% and 30\% for silicates and carbonaceous grains respectively.
Finally, the increase in density also increases the grain destruction level,
though the effect is not nearly as strong as that of decreasing the magnetic
field strength.  The increased density causes the remnant to go radiative at
an earlier time and for shocks that are slower.  The larger cooling rate and
accompanying higher compression caused by the higher density causes more grain
acceleration, however this effect is counteracted to some extent by greater
drag on the grains caused by the higher density.  For the case that we have
explored the increase in gas mass cleared of dust is only 3\% and 7\% for
silicate and carbonaceous grains respectively.  These results are consistent
with those of JTH96 for shocks into a higher density ambient medium.

\section{Discussion}
Our ultimate goal in this work is to understand the evolution of dust in the
ISM.  To make progress in this project we need a way to apply our calculations
of grain processing in individual SNRs to the ISM as a whole.
\citet{Dwek+Scalo_1980} considered these questions and found the following
relation for the dust destruction timescale:
\begin{equation}
\tau_\mathrm{dust} = \frac{\tau_\mathrm{SN} M_\mathrm{ISM}}{M_g} =
\frac{\Sigma_\mathrm{ISM}}{\mathcal{R}_\mathrm{SN} M_g}
\label{eqn:tdust}
\end{equation}
where $\tau_\mathrm{SN}$ is the mean interval between supernovae in the Galaxy
(the inverse of the rate), $M_\mathrm{ISM}$ is the total mass in dust and gas
in the Galaxy, $M_g$ is the mass of gas that is cleared of dust per SN from
Eq.\ \ref{eqn:mdest}, $\Sigma_\mathrm{ISM}$ is the surface density of gas and
dust in the ISM and $\mathcal{R}_\mathrm{SN}$ is the supernova rate per unit
area.  Thus the destruction timescale can be defined in a global way in the
ISM as a whole using $\tau_\mathrm{SN}$ and $M_\mathrm{ISM}$ or locally in some
suitably large region of the ISM using $\Sigma_\mathrm{ISM}$ and
$\mathcal{R}_\mathrm{SN}$. Since the dust destruction timescale depends on the
these combinations of quantities, we will refer to either $\tau_\mathrm{SN}
M_\mathrm{ISM}$ or $\Sigma_\mathrm{ISM}/\mathcal{R}_\mathrm{SN}$ as the SN
mass interval and denote it by $\mathcal{N}$. 

\subsection{SN mass interval}
It is clear that the SN mass interval is a critical factor in assessing the
effectiveness of grain destruction though it depends on two uncertain
quantities. JTH96 and \citet{Bocchio_etal_2014} used the global values from
\citet{McKee_1989},  $\tau_\mathrm{SN} = 125$ yr and $M_\mathrm{ISM} =
4.5\times 10^9 M_\odot$, for a value of $\mathcal{N} = 560$ M$_\odot$ Gyr.
Here $\tau_\mathrm{SN}$ is the ``effective'' SN interval corrected by McKee to
account for the ineffectiveness in destroying dust of SN that go off inside
superbubbles and out of the Galactic plane (some SN Ia's).  \citet{McKee_1989}
estimates that these considerations reduce the effective SN rate by roughly a
factor of 0.36. Recent determinations of the total ISM mass are
substantially higher than the value from \citet{Scoville+Sanders_1987}, which
was used by \citet{McKee_1989}.  For example, \citet{Kalberla+Kerp_2009} give
a value of $12.5\times10^9$ M$_\odot$ with roughly a 15\% uncertainty
(Kalberla, private communication).  A more recent determination of the global
SN rate was carried out by \citet{Li_etal_2011} using a large database of
galaxies in the local universe, and they found a value of $2.84\pm0.60$ SN
(including both types I and II) per century or an interval of 35 yr, which is
a slightly shorter interval than McKee's adopted value of 125 yr, after
including his 36\% correction factor.  Combining these values, without McKee's
correction factor, yields a value of $\mathcal{N} = 440 \pm 120$ M$_\odot$
Gyr.  Including the correction factor this becomes $\mathcal{N} = 1220 \pm
320$ M$_\odot$ Gyr which is significantly larger than found using McKee's
values.

Local estimates of the SN rate and surface density offer the possibility of a
better estimate for the region of the Galaxy near the Sun, for which we have
the best data on the dust content and size distribution, but at the expense of
more uncertainty in the rates themselves.  Estimates of the local values for
$\mathcal{R}_\mathrm{SN}$ and $\Sigma_\mathrm{ISM}$, have recently been made
by \citet{Calura_etal_2010}.  For the SN rate per unit area (combining types I
and II) they use results from  \citet{Cappellaro_1996} and find
$\mathcal{R}_\mathrm{SN} = 0.015\pm0.008$ Gyr$^{-1}\,$pc$^{-2}$.  This rate is
based on the global rate per galaxy for nearby galaxies divided by an area of
$10^9$ pc$^2$ for the Milky Way, making it a global average.  However it is
consistent with the local rate found by \citet{Schmidt_etal_2014} for the
region within 600 pc of the Sun, $\mathcal{R}_\mathrm{SN} =
0.015^{+0.004}_{-0.003}$ Gyr$^{-1}\,$pc$^{-2}$ for core collapse SN.  Global
SN II and SN I estimates \citep[e.g.][]{Li_etal_2011} find that the SN I rate
is about a quarter of the SN II rate, so we increase the
\citet{Schmidt_etal_2014} by 25\%. The estimate of the mass per unit area in
\citet{Calura_etal_2010} is $10.5\pm3.5$ M$_\odot$ pc$^{-2}$, which is a local
surface density using results from \citet{Kulkarni+Heiles_1987},
\citet{Dame_1993} and \citet{Olling+Merrifield_2001}.  Using the rate per unit
area from \citet{Schmidt_etal_2014} (corrected for SN I) and surface density
from \citet{Calura_etal_2010} without the SN rate correction factor yields
$\mathcal{N} = 560\pm 240$ M$_\odot$ Gyr. With the SN rate correction factor
we get $\mathcal{N} = 1560\pm 670$ M$_\odot$ Gyr.

\subsection{Effects of the ISM Model\label{sect:ISM_mod}}
While the formulation of the problem above appears to be correct in a broad
sense, several important factors are not explicitly taken into account.  In
particular, the morphology of the medium, i.e.\ such things as the degree of
homogeneity and the assumed volume filling fractions for the different
phases in the ISM, could strongly affect how SNRs interact with the ISM.  The
temporal and spatial correlation of SN may also be important.  The simplest
assumption is that the medium in which SN explosions occur is completely
uniform in which case the value of $M_g$ in equation \ref{eqn:tdust} is simply
taken directly from Eq.\ \ref{eqn:mdest}.  However if we have a multi-phase
medium, we must consider how the shock propagates in the different phases and
the consequences for grain destruction per SNR.

\begin{figure}[ht!]
\epsscale{0.8}
\plotone{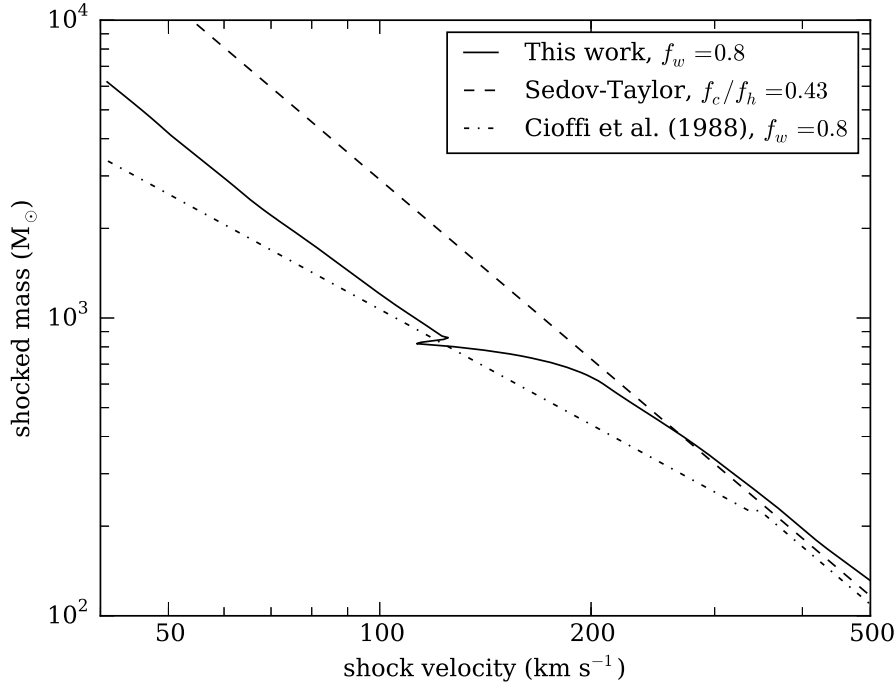}
\caption{Comparison of the mass shocked to a given shock speed for our
hydrodynamical calculations, the Sedov-Taylor evolution used in the hot gas
dominated ISM model of \citet{McKee_1989} and the offset power law
approximation of \citet{Cioffi_etal_1988}. The McKee model (Sedov-Taylor) line
is for an assumed a warm gas cloud filling factor of 0.3 and a hot gas filling
factor of 0.7.  For the results using our hydro calculations and the
\citet{Cioffi_etal_1988} offset power law we assume a warm gas filling factor
of 0.8. (The shocked mass is proportional to $f_w$.) The importance of the
assumed shock expansion model is clear since the amount of dust destroyed is
proportional to mass shocked (see text).}
\label{fig:mshock_vshock}
\end{figure}

JTH96 used the method of \citet{Dwek+Scalo_1980} and \citet{McKee_1989} in
which it is assumed that there is a high filling factor of hot gas and that
the warm gas is confined to clouds embedded in the hot medium. In this picture
the SNR shock expands quickly through the dominant low density hot phase,
remaining adiabatic because of the low density.  When clouds of warm gas are
encountered, they are shocked, though the shock that propagates through the
cloud is at a lower speed because of the higher density.  The post-shock
pressures inside and outside the cloud are roughly equal, which leads to a
shock speed inside the cloud, $v_{sc} \approx \sqrt{\rho_h/\rho_c}\, v_{sh}$,
where the subscript $c$ is for clouds and $h$ is for hot gas.  The mass of
cloud gas shocked to a given shock speed or higher is then given by $M_s = f_c
\rho_c V_s = (f_c/f_h) M_\mathrm{ST}(v_{sc})$, where $M_\mathrm{ST} = 6800
E_{51}/v^2_{s7}$ M$_\odot$ is the mass shocked to at least $v_{s7} = v_s/100$
km s$^{-1}$ in the Sedov-Taylor phase of expansion, $f_c$ is the filling
factor of clouds (warm phase), $f_h$ is the filling factor of hot gas and
$V_s$ is the volume contained in the shock. 
This expression for the mass of gas shocked to a given speed is compared in
Figure \ref{fig:mshock_vshock} with our results for our standard model. The
resulting total mass of gas cleared of dust is then
\begin{equation}
M_g = \frac{f_c}{f_h} 13600 E_{51} \int \frac{\epsilon(v_s)}{v_{s7}^3} \,
dv_{s7}
\label{eqn:Mg_HIM}
\end{equation}
where $\epsilon(v_s)$ is the grain destruction efficiency for a steady shock
with speed $v_s$.  In the figure we have used $E_{51} = 1$ for the
Sedov-Taylor expansion model whereas our standard model uses $E_{51} = 0.5$.
The \citet{Cioffi_etal_1988} model curve uses their equations for the late
time offset power-law expansion and the transition time between early
Sedov-Taylor expansion and power-law expansion. We use $E_{51} = 0.5$ for that
curve. Both the curve for our model result and that for the
\citet{Cioffi_etal_1988} model are multiplied by an assumed filling factor for
warm gas of $f_w = 0.8$ as discussed below. 

Implicit in the above method result is the assumption that the presence of the
clouds does not slow the expansion of the blast wave, which propagates at the
S-T rate.  Also assumed in this formulation is that shocks propagating into
clouds are equivalent, in terms of grain destruction, to plane parallel steady
shocks.  This assumption may be questioned since the shocks are wrapping
around clouds and enveloping them, which could lead to strongly non-steady
effects. 

The picture just described is essentially that of \citet{McKee+Ostriker_1977},
who argued for a large filling factor of hot gas, $\sim 70$\%. The result of a
large hot gas filling factor has been disputed for a number of reasons
\citep[e.g.,][see discussion below]{Slavin+Cox_1992,Ferriere_1998} including
the fact that the hot bubbles created by SNRs are considerably smaller when
the magnetic field is included and that a substantial fraction of SNRs are
spatially and temporally correlated, which decreases their effectiveness in
filling the medium with hot gas.  Though McKee took this latter effect into
account in his estimate of grain destruction, its effect on the hot gas
filling factor was not considered.

Our calculation of grain destruction in an evolving SNR assumes a different
picture from that assumed by JTH96 and \citet{Bocchio_etal_2014} to derive
dust destruction timescales.  Here we have assumed a uniform density,
effectively assuming that $f_w \gg f_h$, where $f_w$ is the filling factor for
the warm medium, no longer assumed to be confined to clouds.  We speculate
that if a remnant runs into a region of hot gas that has a pressure not too
different from that inside the remnant, then the rest of the remnant will
continue to expand as it would in a uniform medium.  In that case, the only
difference would be that $M_g$ in equation \ref{eqn:tdust} would need to be
multiplied by a factor of $f_w$, if we assume that the part of the shock front
that propagates into the hot region does not effectively destroy any dust.
This seems reasonable since the shock from a remnant impinging on a
pre-existing hot bubble will in most cases decay into a sound wave before
encountering a bubble wall. The shocked mass for our hydro calculation plotted
in Figure \ref{fig:mshock_vshock} includes the factor $f_w$ set to 0.8.

\subsection{Volume Filling Factors in the ISM\label{sect:fillfact}}
The volume filling factors for the warm and hot phases of the ISM have yet to
be determined in any definitive way.  \citet{Ferriere_1998} has carried out
probably the most comprehensive study using a range of observational results
and theoretical considerations to arrive at the filling factors of the various
phases of the ISM \citep[see also][]{Ferriere_2001}.  She found that in the
solar neighborhood the warm gas fraction, including both neutral and ionized
phases, is close to 80\%.  In her work as well as that of others, the filling
factors of the different phases vary substantially as a function of
galactic radius and height above the plane.  For the most part, though
in regions near the galactic radius of the Sun and near the Galactic plane,
the filling factor for the hot phase is $\lesssim 20$\%.  Other theoretical
attempts have centered around large scale numerical hydrodynamical or
magneto-hydrodynamical calculations
\citep[e.g.,][]{deAvillez+Breitschwerdt_2005,deAvillez_etal_2012,
Hill_etal_2012}, though achieving sufficient spatial resolution remains
challenging.  Most of these studies in recent years have found modest filling
factors for the hot gas $\lesssim 20$\%, but also for the warm gas, $f_w \sim
20$--50\%.  Often a substantial fraction of the volume is in the thermally
unstable temperature range of $10^4$--$10^{5.5}$ K.  This occurs in these
simulations because of the amount of turbulence injected by SNRs.  

In addition to the resolution issues faced by these simulations, the heating
and cooling rates are often included in very approximate ways.  Since these
rates depend on a range of factors including dust content and the local
ionizing radiation field, the filling factor for warm gas is particularly
difficult to simulate.  Given these considerations, we take as an upper limit
80\% as the filling factor of the warm gas. We assume that there is no
significant grain destruction in the other 20\% of the medium that the SNR
shock overruns, which includes cold, dense gas and hot gas. 

\subsection{Dust Destruction Timescales}
Combining the data on ISM mass and SN rate with modeled SNR evolution and dust
destruction, the dust destruction timescale is
\begin{equation}
\tau_\mathrm{dust} = \frac{\mathcal{N}}{\delta_\mathrm{SN} M_{g,\mathrm{eff}}}
\label{eqn:summary}
\end{equation}
where $\mathcal{N}$ is the SN mass interval, $\delta_\mathrm{SN}$ is the
correction factor intended to account for the effects of correlated SN and SN
out of the plane (which McKee estimates to be 0.36 as discussed above) and
$M_{g,\mathrm{eff}}$ is the effective mass of gas cleared of dust per SN
including the effects of the inhomogeneous ISM.  As discussed above, for the
warm gas dominated ISM, $M_{g,\mathrm{eff}} = f_w M_g$, where $M_g$ is
calculated for an SNR going off in a homogeneous warm medium (see table
\ref{tab:mass_dest}).  For the hot medium \citep[HIM in the terminology
of][]{McKee+Ostriker_1977} dominated case, we can use the methodology of
\citet{Dwek+Scalo_1979} and \citet{McKee_1989} as discussed in \S
\ref{sect:ISM_mod}, with $M_g$ calculated by equation \ref{eqn:Mg_HIM}.  Our
results are summarized in Table \ref{tab:summary}.

The dust destruction timescales listed in the table are substantially higher
than those found by JTH96, 0.37 Gyr for silicate and 0.63 Gyr for cabonaceous
dust.  The factor of 4 to 5 increase is due primarily to the increase of a
factor of $\sim 3$ in the SN mass interval, which in turn derives from a
higher estimate for the ISM mass or surface density.  Another important factor
is the use of a lower SN explosion energy ($5\times10^{51}$ ergs rather than
$10^{51}$ ergs), which lowers the estimate of M$_{g,\mathrm{eff}}$ by roughly
a factor of 2 whether one assumes a HIM dominated medium or a warm gas
dominated medium.  Using the same values as JTH96, including our $E_{51} = 1$
calculation, would lead to dust destruction timescales roughly the same as
those of JTH96.  Nevertheless, while the effective SN explosion energy (after
early cosmic ray losses) is still quite uncertain, it seems clear that the
value from \citet{McKee_1989} for the ISM mass is low.  In addition the values
in Table \ref{tab:summary} use $f_w = 0.8$, which is likely to be an upper
limit.  A lower value for $f_w$ will decrease the effectiveness of SNR grain
destruction, increasing the dust destruction timescale.

\begin{deluxetable}{lrr}
\tablecolumns{3}
\tablewidth{0pc}
\tablecaption{Grain Destruction Timescale Estimates (Gyr)\tablenotemark{a}}
\tablehead{\colhead{ISM model\tablenotemark{b}} & 
\multicolumn{2}{c}{SN mass interval sources}\\
\colhead{} & Local & Global} 
\startdata
\sidehead{Carbonaceous grains}
HIM dominated & $1.6\pm0.7$  & $1.2\pm0.3$  \\
WM dominated  & $3.2\pm1.4$  & $2.6\pm0.7$ \\
\tableline
\sidehead{Silicate grains}
HIM dominated & $0.92\pm0.39$  & $0.72\pm0.20$ \\
WM dominated & $2.0\pm0.8$ & $1.5\pm0.4$  \\
\tableline
\sidehead{SN mass interval\tablenotemark{c} ($\mathcal{N}$ in Gyr M$_\odot$)}
& $560\pm240$ & $440\pm120$ \\
\enddata
\tablenotetext{a}{The timescale estimates depend directly on the SN mass
interval listed in the last line of the table. The errors are from the
estimated errors in the SN mass interval only.} 
\tablenotetext{b}{The HIM dominated values are for the assumption that the
grain destruction occurs in warm clouds embedded in a hot medium as in Eq.\
\ref{eqn:Mg_HIM} and using the plane parallel steady shock model results for
$M_g$ with $f_c/f_h = 0.43$.  The WM dominated values use the hydro results
for our standard model with the M$_g$ calculated as in Eq.\ \ref{eqn:mdest}
multiplied by $f_w = 0.8$.}
\tablenotetext{c}{The SN correction factor, $\delta_\mathrm{SN} = 0.36$, is
not included in the values for $\mathcal{N}$ given in the table but is
included in the timescale values.}
\label{tab:summary}
\end{deluxetable}

For comparison with the destruction timescale, recent estimates for the grain
production timescale center around $2.5\times10^9$ yr
\citep{Jones+Tielens_1994}.  Using the simple estimate of grain equilibrium
fraction of \citet{McKee_1989}, $\delta_\mathrm{eq} = \delta_\mathrm{in} [1 +
t_\mathrm{in}/\tau_\mathrm{dust}]^{-1}$, where $t_\mathrm{in}$ is the dust
injection timescale and $\delta_\mathrm{in}$ is the mass fraction in dust when
injected into the ISM, our standard run destruction timescale values leads to
only about 30--40\% of the silicate elements and 45--50\% of the carbon locked
up in grains in contrast to derived values of $\sim 90$\% and $\sim 50$\% for
the depletion of silicate and carbon grain constituents in the diffuse ISM.
(We have followed JTH96 in assuming $\delta_\mathrm{in} = 0.9$.) Thus with
these results we can explain the carbon depletion fraction but still fall
short of explaining the silicate depletions.  However, with the higher
carbonaceous grain sputtering yields advocated by
\citet{SerraDiaz-Cano+Jones_2008}, the carbonaceous grain destruction
timescales would also be too short.

The discrepancy between the rates of dust production and destruction can also
be illustrated by considering the net effects of SNe on the dust content in
the ISM.  The rate of grain destruction by SNRs depends on the dust-to-gas
mass ratio of the ISM into which the remnant is expanding.  Our results show
that, for a gas-to-dust ratio of 163, a typical SNR destroys $\sim 3.7$
M$_\odot$ of silicate dust and $\sim 1.4$ M$_\odot$ of carbonaceous dust
(using our standard model results).  Core collapse supernovae (CCSN) produce
at most 0.1--0.2 M$_\odot$ of dust
\citep{Gomez_etal_2012,Arendt_etal_2014,Temim_etal_2015}, so if the ISM has
roughly solar metallicity (in gas and dust), SNe are net destroyers of dust.
This discrepancy cannot be made up by dust production in low mass stars.  For
a Salpeter IMF, the number of the most efficient dust producing AGB stars,
with initial masses in the 2--4 M$_\odot$ range, is about 4 times higher than
the number of stars massive enough to become CCSN, but their dust yield is
only $\sim 0.01$ M$_\odot$ \citep{Ventura_etal_2014}.  Thus the relative
ratios between the rates of grain destruction by SNRs, and production by CCSN
and AGB stars is approximately 100:4:1, respectively.  This discrepancy
demands that either a large fraction of the dust is shielded in some way, or
that not all SNe are as efficient in destroying the dust as estimated (e.g.\
because of correlated SNe) or that dust is created in the ISM (e.g.\ via
accretion).  Because the discrepancy is so large, all of these effects likely
come into play. 

\section{Conclusions}
We have evaluated the destruction of interstellar dust grains in supernova
remnants evolving in a homogeneous, magnetized ISM.  We find that the grain
destruction efficiency is not very different from that calculated for steady,
plane parallel shocks with the same shock parameters as the evolving spherical
shocks except for $v_s \gtrsim 200$. Our efficiency calculations have
been carried to lower and higher shock speeds than the previous steady shock
calculations.

Recent grain destruction calculations have used the results of the grain
destruction efficiency in the context of a hot gas dominated ISM 
\citep[JTH96,][]{Bocchio_etal_2014}.  When we use our results self-consistently,
that is assuming that the medium is dominated by warm gas, we find grain
destruction timescales that are larger than those found previously,
e.g. for our standard run with an explosion energy of $5\times10^{50}$ ergs,
using the local estimates of SN rates and mass per unit area yields
$\tau_\mathrm{dust} = 2.0$ Gyr and $3.2$ Gyr for silicates and carbonaceous
grains respectively.  This does not entirely remove the problem of a
destruction timescale that is too short relative to the creation timescale to
account for the level of elemental depletion in the ISM, but does mitigate the
situation. If the volume filling factor of warm ISM is smaller than our
assumed upper limit of 80\% and SNR grain destruction is inefficient in the
other phases of the ISM, then these timescale estimates will increase as
$1/f_w$.  We note that there are many uncertainties that we have not discussed
in this paper concerning both grain destruction and creation as discussed at
length in \citet{Jones+Nuth_2011}. Our results demonstrate the importance of
the morphology of the ISM, in particular the type of medium into which SNRs
typically evolve, for determining dust destruction timescales.

\acknowledgements
This research was supported by NASA Astrophysics Theory Program grant
NNX12AF84G. ED acknowledges support by NASA Astrophysical Data Analysis
Program ADAP13-0094. We thank John Raymond for providing the steady shock
model calculations and our collaborator Xander Tielens for helpful
discussions. We also wish to thank the developers of matplotlib, a python
plotting library, which we used to produce all of the plots in this paper.

\end{document}